\documentclass[10pt,a4paper,twocolumn]{article}
\usepackage{graphicx}

\begin{document}

\title{ A defect of the electromagnetism and it serves as a hidden variable
for quantum mechanics }
\author{Huai-yang Cui \\
Department of physics, Beihang University, Beijing, 100083, China.\\
E-mail: hycui@buaa.edu.cn }
\date{{\small \today}}
\maketitle

\begin{abstract}
{\small According to the theory of relativity, the 4-vector force acting on
a particle is orthogonal to the 4-vector velocity of the particle, but we
found that the electromagnetic force of the Maxwell's theory can not
completely satisfy this orthogonality, this incompletion leads us to find
that the electromagnetic force has two independent components in the 4
dimensional space time. A 4-vector force has 4 components, because of the
orthogonality of 4-vector force and 4-vector velocity, the number of
independent components of the force reduces to 3, while the electromagnetic
force can merely provide 2 independent components, this situation means that
there is an undefined component accompanying the electromagnetic force. This
missing undefined component may serve as a hidden variable which quantum
mechanics is eager to find for a long time. The primary purpose of this
paper is to strictly proof that the electromagnetic force has two
independent components in the 4 dimensional space time. We also briefly
discuss the possibility that the undefined component of the electromagnetic
force serves as a hidden variable for quantum mechanics. }
\end{abstract}
\newline
\newline

\section{ Introduction}

In the theory of relativity, the 4-vector force acting on a particle is
orthogonal to the 4-vector velocity of the particle, because the magnitude
of the 4-vector velocity $u$ keeps constant\cite{Harris} as 
\begin{equation}
|u|=\sqrt{u_{\mu }u_{\mu }}=\sqrt{-c^{2}}=ic  \label{1}
\end{equation}
Any force acting on the particle can never change $u$ in the magnitude but
change $u$ in the direction, thus the 4-vector force is orthogonal to the
4-vector velocity. Strictly, the proofing of the orthogonality is given by 
\begin{equation}
u_{\mu }f_{\mu }=u_{\mu }m\frac{du_{\mu }}{d\tau }=\frac{m}{2}\frac{d(u_{\mu
}u_{\mu })}{d\tau }=0  \label{2}
\end{equation}
The orthogonality does not depend upon any property of 4 vector force, holds
for any kinds of forces. Where the frame of reference is a Cartesian
coordinate system whose axes are orthogonal to one another, there is no
distinction between covariant and contravariant components, only subscripts
need to be used, $m$ denotes the mass of the particle, the 4 dimensional
space time refers to $(x_{1},x_{2},x_{3},x_{4}=ict)$, the index $\mu $ takes
over $1,2,3,4.$

We found that the electromagnetic force of the Maxwell's theory can not
completely satisfy this orthogonality, this incompletion leads us to find
that the electromagnetic force has two independent components in the 4
dimensional space time. A 4-vector force has 4 components, because of the
orthogonality of 4-vector force and 4-vector velocity, the number of
independent components of the force reduces to 3, while the electromagnetic
force can merely provide 2 independent components, this situation means that
there is an undefined component accompanying the electromagnetic force ---
does this missing undefined component is a hidden variable ?--- quantum
mechanics is eager to find it for a long time.

The primary purpose of this paper is to strictly proof that the
electromagnetic force has two independent components in the 4 dimensional
space time. We also briefly discuss the possibility that the undefined
component of the electromagnetic force serves as a hidden variable for
quantum mechanics.

\section{ The relationship between the orthogonality and the Maxwell's
equations}

\subsection{ 4-vector Coulomb's force}

Suppose there are two charged particle $q$ and $q^{\prime}$ locating at the
positions $x$ and $x^{\prime}$ respectively in a Cartesian coordinate system 
$S$, and moving at the 4-vector velocities $u$ and $u^{\prime}$respectively,
as shown in Fig.1, where we have used $X$ to denote $x-x^{\prime}$. The
Coulomb's force $f$ acting on the particle $q$ is orthogonal to the velocity
direction of $q$, as illustrated in Fig.1 by using the Euclidian geometry to
represent the complex space-time, like a centripetal force, the force $f$
would likely make an attempt to rotate itself about the particle's path
center, we think that the path center, $q$, $q^{\prime}$ and the force $f$
all should be in the plane of $u^{\prime}$ and $X$, so that we make an
expansion to $f$ as

\begin{figure}[tbh]
\includegraphics[bb=160 585 350 730,clip]{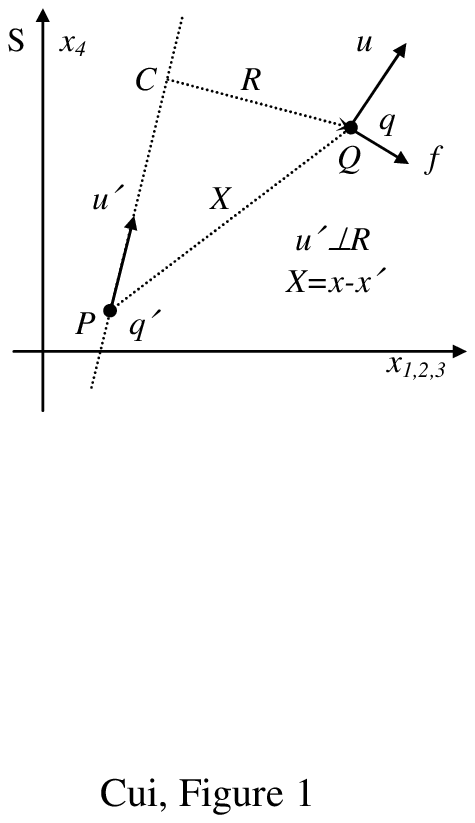}
\caption{The Coulomb's force acting on $q$ is orthogonal to the 4-vector
velocity $u$ of $q$, and lies in the plane of $u^{\prime}$ and $X$ with a
retardation with respect to $q^{\prime }$, here the Euclidian geometry is
used to illustrate the complex space-time.}
\label{Afig1}
\end{figure}

\begin{equation}
f=Au^{\prime }+BX  \label{3}
\end{equation}%
where $A$ and $B$ are unknown coefficients, $u^{\prime }$ and $X$ are chosen
as two independent basis vectors, we will discuss the completion of the
expansion in the subsection 4, but for the moment, the expansion would be
used to clarify a relationship between the orthogonality and the
electromagnetism.

Using the orthogonality $f\perp u$, we get 
\begin{equation}
u\cdot f=A(u\cdot u^{\prime })+B(u\cdot X)=0  \label{4}
\end{equation}%
By eliminating the coefficient $B$, we rewrite Eq.(\ref{3}) as 
\begin{equation}
f=\frac{A}{u\cdot X}[(u\cdot X)u^{\prime }-(u\cdot u^{\prime })X]  \label{5}
\end{equation}%
It follows from the direction of Eq.(\ref{5}) that the unit vector $f^{0}$
of the Coulomb's force is given by 
\begin{equation}
f^{0}=\frac{1}{c^{2}r}[(u\cdot X)u^{\prime }-(u\cdot u^{\prime })X]
\label{6}
\end{equation}%
because of$|f^{0}|=1$, where $r=|R|$, $R\perp u^{\prime }$ as illustrated in
Fig.1. Suppose that the magnitude of the force $f$ has the classical form 
\begin{equation}
|f|=k\frac{qq^{\prime }}{r^{2}}  \label{9}
\end{equation}%
Combination of Eq.(\ref{6}) with Eq.(\ref{9}), we obtain a modified
Coulomb's force 
\begin{eqnarray}
f &=&\frac{kqq^{\prime }}{c^{2}r^{3}}[(u\cdot X)u^{\prime }-(u\cdot
u^{\prime })X]  \nonumber \\
&=&\frac{kqq^{\prime }}{c^{2}r^{3}}[(u\cdot R)u^{\prime }-(u\cdot u^{\prime
})R]  \nonumber \\
&=&q[(u\cdot (\frac{kq^{\prime }}{c^{2}r^{3}}R))u^{\prime }-(u\cdot
u^{\prime })(\frac{kq^{\prime }}{c^{2}r^{3}}R)]  \label{10}
\end{eqnarray}%
Using the relation 
\begin{equation}
\partial _{\mu }(\frac{1}{r})=-\frac{R_{\mu }}{r^{3}}  \label{11}
\end{equation}%
we obtain 
\begin{eqnarray}
f_{\mu } &=&q[-(u_{\nu }\partial _{\nu }(\frac{kq^{\prime }}{c^{2}r}))u_{\mu
}^{\prime }+(u_{\nu }u_{\nu }^{\prime })\partial _{\mu }(\frac{kq^{\prime }}{%
c^{2}r})]  \nonumber \\
&=&q[-(u_{\nu }\partial _{\nu }(\frac{kq^{\prime }u_{\mu }^{\prime }}{c^{2}r}%
))+(u_{\nu })\partial _{\mu }(\frac{kq^{\prime }u_{\nu }^{\prime }}{c^{2}r})]
\label{12}
\end{eqnarray}%
The force can be rewritten in terms of 4-vector components as 
\begin{eqnarray}
A_{\mu } &=&\frac{kq^{\prime }u_{\mu }^{\prime }}{c^{2}r}  \label{13} \\
F_{\mu \nu } &=&\partial _{\mu }A_{\nu }-\partial _{\nu }A_{\mu }  \label{14}
\\
f_{\mu } &=&qF_{\mu \nu }u_{\nu }  \label{15}
\end{eqnarray}%
Thus $A_{\mu }$ expresses the 4-vector potential of the particle $q^{\prime
} $. It is easy to find that Eq.(\ref{15}) contains the Lorentz force.

\subsection{ The Lorentz gauge condition and the Maxwell's equations}

From Eq.(\ref{13}), because of $u^{\prime }\perp R$ , i.e. $u_{\mu }^{\prime
}R_{\mu }=0$, we have 
\begin{equation}
\partial _{\mu }A_{\mu }=\frac{kq^{\prime }u_{\mu }^{\prime }}{c^{2}}%
\partial _{\mu }(\frac{1}{r})=-\frac{kq^{\prime }u_{\mu }^{\prime }}{c^{2}}(%
\frac{R_{\mu }}{r^{3}})=0  \label{16}
\end{equation}%
It is known as the Lorentz gauge condition. To note that $R$ has three
degrees of freedom under the condition $R\perp u^{\prime }$, so we have 
\begin{eqnarray}
\partial _{\mu }R_{\mu } &=&3  \label{17} \\
\partial _{\mu }\partial _{\mu }(\frac{1}{r}) &=&-4\pi \delta (R)  \label{18}
\end{eqnarray}%
From Eq.(\ref{14}), we have 
\begin{eqnarray}
\partial _{\nu }F_{\mu \nu } &=&\partial _{\nu }\partial _{\mu }A_{\nu
}-\partial _{\nu }\partial _{\nu }A_{\mu }=-\partial _{\nu }\partial _{\nu
}A_{\mu }  \nonumber \\
&=&-\frac{kq^{\prime }u_{\mu }^{\prime }}{c^{2}}\partial _{\nu }\partial
_{\nu }(\frac{1}{r})=\frac{kq^{\prime }u_{\mu }^{\prime }}{c^{2}}4\pi \delta
(R)  \nonumber \\
&=&\mu _{0}J_{\nu }^{\prime }  \label{19}
\end{eqnarray}%
where we define $J_{\nu }^{\prime }=q^{\prime }u_{\nu }^{\prime }\delta (R)$
as the current density of the source. From Eq.(\ref{14}), by exchanging the
indices and taking the summation of them, we have 
\begin{equation}
\partial _{\lambda }F_{\mu \nu }+\partial _{\mu }F_{\nu \lambda }+\partial
_{\nu }F_{\lambda \mu }=0  \label{20}
\end{equation}%
The Eq.(\ref{19}) and (\ref{20}) are known as the Maxwell's equations. For
continuous media, they are valid as well as.

\subsection{ The Lienard-Wiechert potential}

From the Maxwell's equations, we know that there is a retardation time for
the electromagnetic action to propagate between the two particles, as
illustrated in Fig.1, the retardation effect is measured by 
\begin{equation}
r=c\Delta t=c\frac{|PC|}{ic}=c\frac{u^{\prime 0}\cdot X}{ic}=\frac{%
u^{\prime}_{\nu }(x^{\prime}_{\nu }-x_{\nu })}{c}  \label{21}
\end{equation}
wherer $u^{\prime 0}$ is the unit vecotr of $u^{\prime}$.Then 
\begin{equation}
A_{\mu }=\frac{kq^{\prime}}{c^{2}}\frac{u^{\prime}_{\mu }}{r}=\frac{%
kq^{\prime}}{c}\frac{u^{\prime}_{\mu }}{u^{\prime}_{\nu }(x^{\prime}_{\nu
}-x_{\nu })}  \label{22}
\end{equation}
Obviously, Eq.(\ref{22}) is known as the Lienard-Wiechert potential for a
moving particle.

\subsection{ The completion of the electromagnetic force's basis vectors}

The above formalism clearly shows that the Maxwell's equations can be
derived from the classical Coulomb's force and the orthogonality of 4-vector
force and 4-vector velocity. In other words, the orthogonality is hidden in
the Maxwell's equation. Specially, Eq.(\ref{5}) directly accounts for the
geometrical meanings of the curl of vector potential, the curl contains the
orthogonality. The orthogonality of 4-vector force and 4-vector velocity is
one of consequences from the relativistic mechanics\cite{Cui01}.

But, whether the expansion we have used in Eq.(\ref{3}) is complete?
Obviously, we choose two basis vector $u^{\prime}$ and $X$ to construct the
electromagnetism successfully, this choice means that the electromagnetic
force has two independent components in its Hilbert space based on the two
basis vector $u^{\prime}$ and $X$, we readily recognize that the expansion
is incomplete. A 4-vector force has 4 components, because of the
orthogonality of 4-vecor force and 4-vector velocity, the number of
independent components of the force reduces to 3, thus a complete expansion
of a 4-vector force needs 3 basis vectors.

Our argument also arises from that if the electromagnetic force of the
Maxwell's theory has two independent components, then there is an undefined
component accompanying the Maxwell's electromagnetic force. This missing
undefined component may serve a hidden variable for quantum mechanics. In
the next sections, we will strictly and comprehensively proof that the
electromagnetic force of the Maxwell's theory does have two independent
components, does have an undefined component in the physics. In order to
clarify its aspects as much as possible, we present four methods to proof
the theme.

\section{ The proofs that the electromagnetic force of the Maxwell's theory
has two independent components}

\subsection{ the first proof --- by an intuitive method}

In usual 3D space, the vector product of two vector $\mathbf{a}$ and $%
\mathbf{b}$ is defined as another vector $\mathbf{c}$ given by $\mathbf{c}=%
\mathbf{a}\times \mathbf{b}$, or $c_{k}=\varepsilon _{kij}a_{i}b_{j}$, where
the indexes i, j and k take over $1,2,3$ in the 3D space, $\varepsilon _{kij}
$ defined in usual textbooks; as we know,the vector $\mathbf{c}$ is
orthogonal to both the vectors $\mathbf{a}$ and $\mathbf{b}$, i.e. $\mathbf{c%
}\perp \mathbf{a}$ and $\mathbf{c}\perp \mathbf{b}$. In the Minkowski's
space, we also intruduce a definition for vector product: as shown in Fig.1,
imaging that there is a 4-vector $\Gamma $ which is orthogonal to the plain
of $u^{\prime }$ and $X$, then $\Gamma $ is defined by the vector product as 
\begin{equation}
\Gamma =u^{\prime }\times X\qquad \qquad \Gamma _{\lambda }=\varepsilon
_{\lambda \mu \nu }u_{\mu }^{\prime }X_{\nu }  \label{23}
\end{equation}%
where the indexes take 1,2,3,4. It is easy to obtain their orthogonalities $%
\Gamma \perp u^{\prime }$ and $\Gamma \perp X$, because 
\begin{eqnarray}
\Gamma \cdot X &=&\Gamma _{\lambda }X_{\lambda }=\varepsilon _{\lambda \mu
\nu }u_{\mu }^{\prime }X_{\nu }X_{\lambda }  \nonumber \\
&=&\varepsilon _{\lambda \mu \nu }u_{\mu }^{\prime }X_{\nu }X_{\lambda
}|_{\nu <\lambda }+\varepsilon _{\lambda \mu \nu }u_{\mu }^{\prime }X_{\nu
}X_{\lambda }|_{\nu >\lambda }  \nonumber \\
&=&\varepsilon _{\lambda \mu \nu }u_{\mu }^{\prime }X_{\nu }X_{\lambda
}|_{\nu <\lambda }+\varepsilon _{\nu \mu \lambda }u_{\mu }^{\prime
}X_{\lambda }X_{\nu }|_{\nu <\lambda }  \nonumber \\
&=&(\varepsilon _{\lambda \mu \nu }+\varepsilon _{\nu \mu \lambda })u_{\mu
}^{\prime }X_{\nu }X_{\lambda }|_{\nu <\lambda }=0  \label{24} \\
\Gamma \cdot u^{\prime } &=&0  \label{25}
\end{eqnarray}%
The vector $\Gamma $ can be easily imaged out intuitively in Fig.1. The
electromagnetic 4-vector force has 4 components, it subjects to the first
constraint given by 
\begin{equation}
f\cdot u=0  \label{26}
\end{equation}%
the second constraint on the 4-vector force is given by 
\begin{equation}
f\cdot \Gamma =(Au^{\prime }+BX)\cdot \Gamma =0  \label{27}
\end{equation}%
The first constraint is just the orthogonality of 4-vector force and
4-vector velocity, the second constraint reflects our choice that the
electromagnetic force lies in the plain of $u^{\prime }$ and $X$ to which
the vector $\Gamma $ is orthogonal. Subjecting to these two constraints, the
number of independent components of the electromagnetic force reduces to 2.
The two independent components allow the electromagnetic force to change in
the plain of $u^{\prime }$ and $X$.

\subsection{ the second proof --- in its Hilbert space}

The moving particle $q^{\prime}$ exerts an electromagnetic force $f$ on
another moving particle $q$, the action has been sophisticatedly expressed
as 
\begin{equation}
f_{\mu }=qF_{\mu \nu }u_{\nu }=qu_{\nu }\partial _{\mu }A_{\nu }-qu_{\nu
}\partial _{\nu }A_{\mu }  \label{28}
\end{equation}
To note that in the present stage of human's knowledge, the above each term
can not be resolved into more details, therefore, we conclude that the
electromagnetic force contains two terms, each term represents a basis
vector of its Hilbert space, the number of independent components of the
electromagnetic force is 2.

This provides an insight into electromagnetism that a full electromagnetic
force should contain at least tree terms in its formalism.

\subsection{ the third proof --- hydrogen atom as an instance}

The nucleus of a hydrogen atom provides a spherical Coulomb's electric
potential for its outer electron, this 4-vector force subjects to the first
and second constraints, given by 
\begin{eqnarray}
f\cdot u &=&0  \label{29} \\
\mathbf{f}\times \mathbf{r} &=&0  \label{30}
\end{eqnarray}%
The second constraint expresses that the torque of the centripetal force is
zero, and it arises from the above mentioned $f\cdot \Gamma =0$ for this
instance. Because, refering to Fig.1, $f\cdot \Gamma =f\cdot (u^{\prime
}\times X)\mathbf{=}f\cdot (u^{\prime }\times R)\mathbf{=}0$ permits the
expansion of $f=Au^{\prime }+BR$, whereas the nucleus is at rest, $u^{\prime
}=(0,0,0,ic)$, $R=(\mathbf{r},0)$,  then $f=(B\mathbf{r},Aic)$ i.e., $%
\mathbf{f\parallel r}$ or  $\mathbf{f}\times \mathbf{r}=0$. Inversely, if $%
\mathbf{f\parallel r}$, then we can construct a vector $\Gamma $ so that $%
f\cdot \Gamma =0$. You can see, if we consider the two constraints, then the
4-force in hydrogen atom has only two independent components.

In Bohr's hydrogen atom, the electron of hydrogen moves in a circular 
orbit, the attractive force subjects to the spheric symmetry as its third
constraint, in that case, the number of independent components of the
4-vector force reduces to 1, in consistency with its circular motion.

\subsection{ the fourth proof --- outside the electromagnetism}

A 4-vector force has 4 components, because of the orthogonality of 4-vector
force and 4-vector velocity, the number of independent components reduces to
3, while the electromagnetic force can merely provide 2 independent
components, this situation means that there is an undefined component
accompanying the electromagnetic force. what is the effect of the
electromagnetism-undefined-component (EUC) in the physics? Outside the
electromagnetism, a variety of problems may concern with the EUC, for
example, quantum mechanics is eager to find a hidden variable for a long
time for which the EUC may serve. Since its complex, we put the fourth proof
in the next section, where we choose and briefly discuss three basic
questions: (1) the properties of the EUC; (2) wave function and 4-vector
velocity field in the EUC ensemble space; (3) a successful application to
hydrogen's fine structure.

\section{ The proof outside the electromagnetism}

\subsection{ the properties of the electromagnetism undefined component (EUC)%
}

The properties of the EUC include : (1) EUC is a small quantity which is
hardly noted by physicists, if it serves as a hidden variable for quantum
mechanics, then it should be responsible to the Planck's constant; (2) EUC
is a fluctuating quantity in the ensemble space of enormous identical
experiments, its average value at a given point must be zero, like $%
<f_{EUC}>=0$; (3)as a nature of independency, EUC must be independent from
our electromagnetic field, for example, in a vacuum when any our
electromagnetic field has vanished off its mater environment, while the EUC
can still alive in the vacuum and cause the indeterminacy of the motion of
particles moving in the vacuum; (4)EUC also correlates with our
electromagnetic field through the orthogonality $f\cdot u=0$; (5) The effect
of EUC on physics must be comprehensive, it is not concern with individual
cases, no body can really escape from its fluctuation if the EUC still
remains at undefined status.

According to the above mentioned properties of EUC, the proof outside the
electromagnetism will carry out on "the path" that leads us from the EUC to
the two key points of quantum mechanics: wave function and hydrogen atom.

\subsection{ wave function and 4-vector velocity field in ensemble space}

Consider a particle diffraction experiment in an electromagnetic field where
the EUC is active and cause the indeterminacy of quantum mechanics, Each
particle in the field subjects to both the electromagnetic force $f$ and EUC 
$f_{EUC}$, then it satisfies the dynamic equation: 
\begin{eqnarray}
m\frac{du_{\mu }}{d\tau } &=&f_{\mu }+f_{EUC\mu }=qF_{\mu \nu }u_{\nu
}+f_{EUC\mu }  \label{32} \\
u_{\mu }u_{\mu } &=&-c^{2}  \label{33}
\end{eqnarray}%
In order to eliminate the EUC term by means of statistics, i.e. using $%
<f_{EUC}>=0$, we turn to study this dynamic equation in the ensemble space
which consists of enormous particle paths recorded in enormous identical
experiments, in the ensemble space we find 
\begin{eqnarray}
m\frac{d<u_{\mu }>}{d\tau } &=&qF_{\mu \nu }<u_{\nu }>  \label{34} \\
<u_{\mu }><u_{\mu }> &=&-c^{2}  \label{35}
\end{eqnarray}%
Strictly speaking, Eq.(\ref{35}) is not derived from Eq.(\ref{33}), it
stands for the property that the magnitude of any 4-vector velocity keeps
constant, otherwise this velocity concept collapses. If we regard the
enormous recorded particle paths in the ensemble space as a flow, it is easy
to find that there a 4-vector velocity field for the flow. We clearly
emphasize two points:(1) At every point of the ensemble space, the mean
4-vector velocity satisfies the above mean dynamic equation; (2) the mean
4-velocity is a 4-vector velocity field of the flow, $<u>$ is a function of
the position of the ensemble space, i.e. in the form of $%
<u>(x_{1},x_{2},x_{3},x_{4})$.

For our convenience, we drop the mean sign $<>$ for the 4-vector velocity $u$
in the followings, simply use $u$ in palce of $<u>$. As mentioned above, the
4-vector velocity $u$ can be regarded as a 4-vector velocity field in the
ensemble space, then 
\begin{eqnarray}
\frac{du_{\mu }}{d\tau } &=&\frac{\partial u_{\mu }}{\partial x_{\nu }}\frac{%
dx_{\nu }}{d\tau }=u_{\nu }\partial _{\nu }u_{\mu }  \label{1a} \\
qF_{\mu \nu }u_{\nu } &=&qu_{\nu }(\partial _{\mu }A_{\nu }-\partial _{\nu
}A_{\mu })  \label{1b}
\end{eqnarray}%
Beware: Eq.(\ref{1a}) is the most important step in the present work after
inducing the concept of 4-vector velocity field in ensemble space.
Substituting them back into the dynamic equation, and re-arranging these
terms, we obtain 
\begin{eqnarray}
u_{\nu }\partial _{\nu }(mu_{\mu }+qA_{\mu }) &=&u_{\nu }\partial _{\mu
}(qA_{\nu })  \nonumber \\
&=&u_{\nu }\partial _{\mu }(mu_{\nu }+qA_{\nu })-u_{\nu }\partial _{\mu
}(mu_{\nu })  \nonumber \\
&=&u_{\nu }\partial _{\mu }(mu_{\nu }+qA_{\nu })-\frac{1}{2}\partial _{\mu
}(mu_{\nu }u_{\nu })  \nonumber \\
&=&u_{\nu }\partial _{\mu }(mu_{\nu }+qA_{\nu })-\frac{1}{2}\partial _{\mu
}(-mc^{2})  \nonumber \\
&=&u_{\nu }\partial _{\mu }(mu_{\nu }+qA_{\nu })  \label{1c}
\end{eqnarray}%
Using the notation 
\begin{equation}
K_{\mu \nu }=\partial _{\mu }(mu_{\nu }+qA_{\nu })-\partial _{\nu }(mu_{\mu
}+qA_{\mu })  \label{1d}
\end{equation}%
Eq.(\ref{1c}) is given by 
\begin{equation}
u_{\nu }K_{\mu \nu }=0  \label{1f}
\end{equation}%
Because $K_{\mu \nu }$ contains the variables $\partial _{\mu }u_{\nu }$, $%
\partial _{\mu }A_{\nu }$, $\partial _{\nu }u_{\mu }$ and $\partial _{\nu
}A_{\mu }$, they are independent from $u_{\nu }$, then a solution satisfying
Eq.(\ref{1f}) is actually of

\begin{eqnarray}
K_{\mu \nu } &=&0  \label{1g} \\
\partial _{\mu }(mu_{\nu }+qA_{\nu }) &=&\partial _{\nu }(mu_{\mu }+qA_{\mu
})  \label{1h}
\end{eqnarray}%
The above equation allows us to introduce a potential function $\Phi $ in
mathematics, further set $\Phi =-i\hbar \ln \psi $, we obtain a very
important equation 
\begin{equation}
(mu_{\mu }+qA_{\mu })\psi =-i\hbar \partial _{\mu }\psi  \label{2a}
\end{equation}%
where $\psi $ may be a complex mathematical function, its physical meanings
will be determined from experiments after the introduction of the Planck's
constant $\hbar $, as we have know, it is wave function.

Substituting Eq.(\ref{2a}) into $u_{\mu }u_{\mu }=-c^{2}$, we obtain a wave
equation 
\begin{equation}
(-i\hbar \partial _{\mu }\psi -qA_{\mu }\psi )(-i\hbar \partial _{\mu }\psi
-qA_{\mu }\psi )=-m^{2}c^{2}\psi ^{2}  \label{3a}
\end{equation}%
It is a new quantum wave equation\cite{Cui02}. Where the left side
corresponds to the product of momentum and momentum itself, does not
correspond to the product of momentum operator and momentum operator.

In this subsection, by using the $<f_{EUC}>=0$ concept and 4-vector velocity
field in ensemble space, we construct a relation between wave function and
particle momentum, as we know, the relation has been widely used in quantum
mechanics.

\subsection{ an application to hydrogen atom's fine structure}

In the following, we use the Gaussian units, and use $m_{e}$ to denote the
rest mass of electron. In a spherical polar coordinate system $(r,\theta
,\varphi ,ict)$, the nucleus of a hydrogen atom provides a spherically
symmetric potential $V(r)=e/r$ for the electron motion. The wave equation (%
\ref{3a}) for the hydrogen atom in the energy eigenstate $\psi (r,\theta
,\varphi )e^{-iEt/\hbar }$ may be written in the spherical coordinates: 
\begin{eqnarray}
\frac{m_{e}^{2}c^{2}}{\hbar ^{2}}\psi ^{2} &=&(\frac{\partial \psi }{%
\partial r})^{2}+(\frac{1}{r}\frac{\partial \psi }{\partial \theta })^{2}+(%
\frac{1}{r\sin \theta }\frac{\partial \psi }{\partial \varphi })^{2} 
\nonumber \\
&&+\frac{1}{\hbar ^{2}c^{2}}(E+\frac{e^{2}}{r})^{2}\psi ^{2}  \label{4a}
\end{eqnarray}%
By substituting $\psi =R(r)X(\theta )\phi (\varphi )$, we separate the above
equation into 
\begin{eqnarray}
(\frac{\partial \phi }{\partial \varphi })^{2}+\kappa \phi ^{2} &=&0 
\nonumber \\
&&  \label{5a} \\
(\frac{\partial X}{\partial \theta })^{2}+[\lambda -\frac{\kappa }{\sin
^{2}\theta }]X^{2} &=&0  \nonumber \\
&&  \label{6a} \\
(\frac{\partial R}{\partial r})^{2}+[\frac{1}{\hbar ^{2}c^{2}}(E+\frac{e^{2}%
}{r})^{2}-\frac{m_{e}^{2}c^{2}}{\hbar ^{2}}-\frac{\lambda }{r^{2}}]R^{2} &=&0
\nonumber \\
&&  \label{7a}
\end{eqnarray}%
where $\kappa $ and $\lambda $ are constants introduced for the separation.
Eq.(\ref{5a}) can be solved immediately, with the requirement that $\phi
(\varphi )$ must be a periodic function, we find its solution given by 
\begin{equation}
\phi =C_{1}e^{\pm i\sqrt{\kappa }\varphi }=C_{1}e^{-im\varphi } \quad m=\pm 
\sqrt{\kappa }=0,\pm 1,...  \label{8a}
\end{equation}%
where $C_{1}$ is an integral constant. It is easy to find the solution of
Eq.(\ref{6a}), it is given by 
\begin{equation}
X(\theta )=C_{2}\exp ( \pm i\int \sqrt{\lambda -\frac{m^{2}}{\sin ^{2}\theta 
}}d\theta )  \label{9a}
\end{equation}%
where $C_{2}$ is an integral constant. The requirement of periodic function
for $X$ demands 
\begin{equation}
\int_{0}^{2\pi }\sqrt{\lambda -\frac{m^{2}}{\sin ^{2}\theta }}d\theta =2\pi
k \quad k=0,1,2,...  \label{10a}
\end{equation}%
This complex integration has been evaluated (see the appendix of the ref.%
\cite{Cui04}, we get 
\begin{equation}
\int_{0}^{2\pi }\sqrt{\lambda -\frac{m^{2}}{\sin ^{2}\theta }}d\theta =2\pi (%
\sqrt{\lambda }-|m|)  \label{11a}
\end{equation}%
thus, we obtain 
\begin{equation}
\sqrt{\lambda }=k+|m|  \label{12a}
\end{equation}%
We rename the integer $\lambda $ as $j^{2}$ for a convenience in the
followings, i.e. $\lambda =j^{2}$. The solution of Eq.(\ref{7a}) is given by 
\begin{equation}
R(r)=C_{3}\exp (\pm \frac{i}{\hbar c}\int \sqrt{(E+\frac{e^{2}}{r}%
)^{2}-m_{e}^{2}c^{4}-\frac{\lambda \hbar ^{2}c^{2}}{r^{2}}}dr)  \label{13a}
\end{equation}%
where $C_{3}$ is an integral constant. The requirement that the radical wave
function forms a "standing wave" in the range from $r=0$ to $r=\infty $
demands

\begin{eqnarray}
&&\frac{1}{\hbar c}\int_{0}^{\infty }\sqrt{(E+\frac{e^{2}}{r}%
)^{2}-m_{e}^{2}c^{4}-\frac{\lambda \hbar ^{2}c^{2}}{r^{2}}}dr  \nonumber \\
&=&\pi s \quad s=0,1,2,...  \label{14a}
\end{eqnarray}%
This complex integration has been evaluated (see the appendix of the ref.%
\cite{Cui04}), we get

\begin{eqnarray}
&&\frac{1}{\hbar c}\int_{0}^{\infty }\sqrt{(E+\frac{e^{2}}{r}%
)^{2}-m_{e}^{2}c^{4}-\frac{j^{2}\hbar ^{2}c^{2}}{r^{2}}}dr  \nonumber \\
&=&\frac{\pi E\alpha }{\sqrt{m_{e}^{2}c^{4}-E^{2}}}-\pi \sqrt{j^{2}-\alpha
^{2}}  \label{15a}
\end{eqnarray}%
where $\alpha =e^{2}/\hbar c$ is known as the fine structure constant.

From the last Eq.(\ref{14a}) and Eq.(\ref{15a}), we obtain the energy levels
given by 
\begin{equation}
E=m_{e}c^{2}[1+\frac{\alpha ^{2}}{(\sqrt{j^{2}-\alpha ^{2}}+s)^{2}}]^{-1/2}
\label{16a}
\end{equation}%
where $j=\sqrt{\lambda }=k+|m|$. Because the restriction $j\neq 0$ in Eq.(%
\ref{16a}), we find $j=1,2,3...$.

The result, Eq.(\ref{16a}), is completely the same as the calculation of the
Dirac wave equation for the hydrogen atom\cite{Schiff}, it is just the 
\textbf{fine structure of hydrogen atom energy}.

After we introduce the EUC into quantum mechanics, the quantum mechanics
becomes reasonable at its key points, we think that we have indirectly
proofed the existence of the EUC outside the electromagnetism, and the
electromagnetic force of the Maxwell's theory has two independent components
in the 4 dimensional space-time.

\section{ Conclusions}

In this paper, we proof that the electromagnetic force of the Maxwell's
theory has two independent components in the 4 dimensional space time. A
4-vector force has 4 components, because of the orthogonality of 4-vector
force and 4-vector velocity, the number of independent components of the
force reduces to 3, while the electromagnetic force can merely provide 2
independent components, this situation means that there is an undefined
component accompanying the electromagnetic force. Therefore, we also briefly
and confirmedly discuss the possibility that the undefined component of the
electromagnetic force serves as a hidden variable for quantum mechanics.


\begin{thebibliography}{9}
\bibitem{Harris} E. G. Harris, Introduction to Modern Theoretical Physics,
Vol.1, p.263, Eq.10-40, John Wiley \&Sons, USA, 1975.

\bibitem{Cui01} H. Y. Cui, "The direction of Coulomb's force and the
direction of gravitational force in the 4-dimensional space time",
arXiv:physics/0102073, Oct, 2001.

\bibitem{Cui02} H. Y. Cui, "The calculation of vortex structures in
superconductivity", in "The proceedings of the annual meeting of the Chinese
physical society, 2002", Aug., 2002.

\bibitem{Cui04} H. Y. Cui, "The novel aspect of hydrogen atom: the fine
structure and spin effect can be derived by single component wavefunction",
arXiv:physics/0408025, Aug., 2004.

\bibitem{Schiff} L. I. Schiff, Quantum Mechanics, third ed., p.486,
Eq.53.26, McGrall-Hill, USA, 1968.
\end{thebibliography}
\end{document}